\documentclass[11pt,twoside]{article}

\usepackage{asp2006}
\usepackage{epsf}
\usepackage{psfig}
\usepackage{lscape}

\begin{document}
\title{Jet-environment interactions in FRI radio galaxies}   
\author{R.A. Laing}   
\affil{ESO, Karl-Schwarzschild-Stra\ss e 2, 85748 Garching-bei-M\"{u}nchen,
  Germany}    
\author{A.H. Bridle}   
\affil{NRAO, 520 Edgemont Road, Charlottesville, VA
22903-2475, U.S.A.}    

\begin{abstract} 
There is now unequivocal evidence that the jets in FR I radio galaxies are
initially relativistic, decelerating flows. On the assumption that they are
axisymmetric and intrinsically symmetrical (a good approximation close to the
nucleus), we can make models of their geometry, velocity, emissivity and field
structure whose parameters can be determined by fitting to deep VLA
observations. Mass entrainment -- either from stellar mass loss within the jet
volume or via a boundary layer at the jet surface -- is the most likely cause
for deceleration. This idea is quantitatively consistent with the velocity field
and geometry inferred from kinematic modelling and the external gas density and
pressure profiles derived from X-ray observations.  The jets must initially be
very light, perhaps with an electron-positron composition.
\end{abstract}



\section{Introduction}

\begin{figure}[!ht]
\plotone{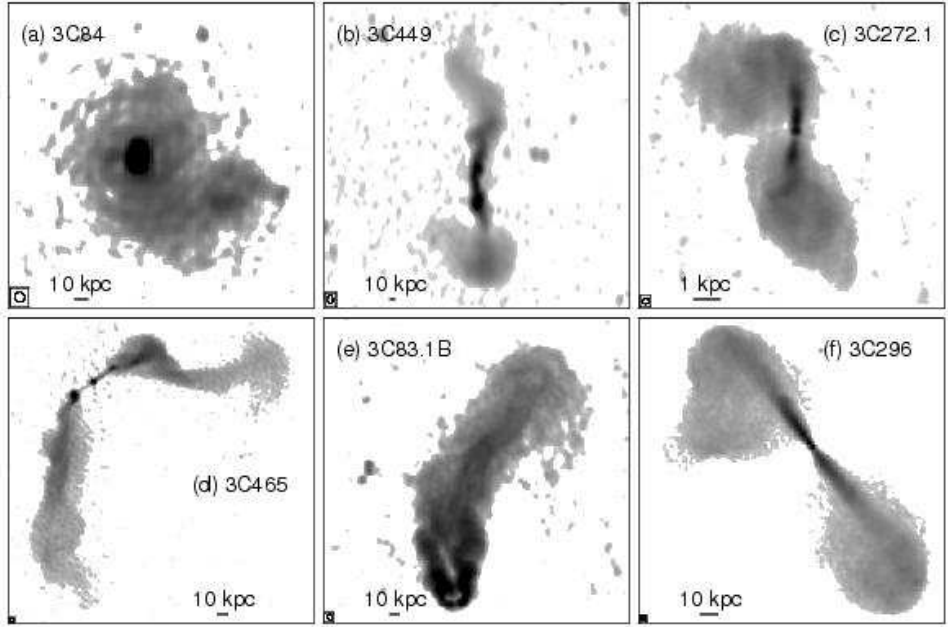}
\caption{Examples of the radio structures of FR\,I sources from the 3CRR
  catalogue \citep{3CRRAtlas}.\label{fig:fris}}
\end{figure}

The morphological division of extragalactic radio sources into two classes 
introduced by \cite{FR74} has proved to be remarkably robust. Fanaroff \& Riley
had already noted that the edge-brightened (FR\,II) sources tend to have higher
radio luminosities than the edge-darkened (FR\,I) sources, but there is also a
dependence on stellar luminosity of the host galaxy which makes the division
even cleaner \citep{Ledlow}.  That said, the FR\,I class is not homogeneous:
examples of the range of structures are shown in
Fig.~\ref{fig:fris}. Although almost all FR\,I sources show jets on small
scales, some appear to be confined to the nuclear regions (e.g.\ 3C\,84;
Fig.~\ref{fig:fris}a) while others persist almost to the outer edges of the
source structure (e.g.\ 3C\,296; Fig.~\ref{fig:fris}f). 

It has long been surmised (e.g.\ \citealt{Simon78,Fanti82}) that the
characteristic structures of FR\,I sources result from deceleration and
disruption of their jets, perhaps triggered by interaction with the surrounding
intergalactic medium. Turning this perception from a vague idea into a
quantitative description of source dynamics proved to be a slow process, but
significant progress has recently been made using a combination of deep radio
imaging, sophisticated jet modelling and an understanding of the external
environments of the sources from X-ray observations.  These results bear on five
of \citet{Blandford}'s tasks: map jet velocity fields; understand the changing
composition; measure jet pressures; deduce jet confinement mechanisms and infer
jet powers and thrusts.

\section{FR\,I jets as decelerating, relativistic flows}

\begin{figure}[!ht]
\plotone{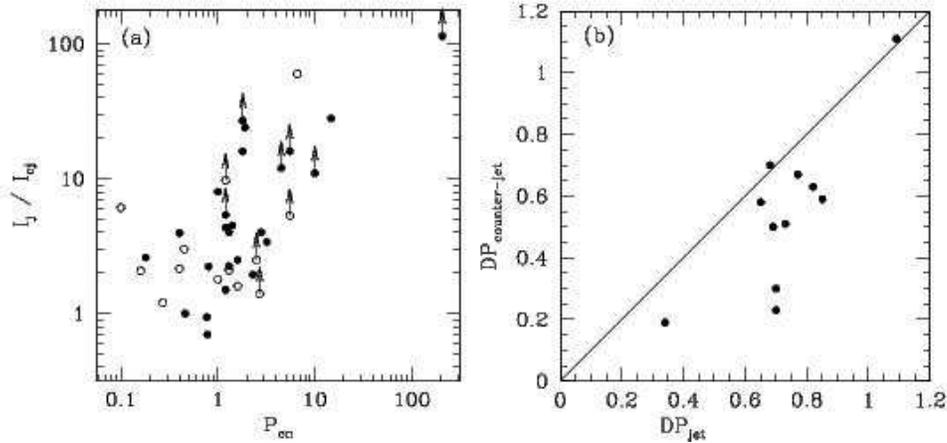}
\caption{(a) A plot of jet/counter-jet intensity ratio at the brightening point,
  $I_{\rm j}/I_{\rm cj}$, against normalized core power, P$_{\rm CN}$ for
  sources from the B2 sample \citep{LPdRF}. (b) Depolarization on the
  counter-jet (DP$_{\rm counter-jet}$) and jet (DP$_{\rm jet}$) sides for
  sources from the same sample with $I_{\rm j}/I_{\rm cj} > 4$
  \citep{Morganti97}. \label{fig:relclues}}
\end{figure}

Deceleration of a symmetric, initially relativistic flow provides a very natural
explanation for the decrease in jet/counter-jet sidedness ratio $I_{\rm
j}/I_{\rm cj}$ with distance from the nucleus observed in FR\,I jets
(e.g.\ \citealt{Lai93}).  There is now direct evidence for relativistic
motions of jet knots on kpc scales in M\,87 and Cen\,A
\citep{BZO95,Harris07,Hard03} and many other results are mostly easily explained
if FR\,I jet bases are relativistic. For example, $I_{\rm j}/I_{\rm cj}$ is
correlated with core fraction \citep[Fig.~\ref{fig:relclues}a]{LPdRF} and the
counter-jet lobe depolarizes more rapidly with increasing wavelength than that
associated with the brighter jet \citep[Fig.~\ref{fig:relclues}b]{Morganti97}.

\begin{figure}[!ht]
\plotfiddle{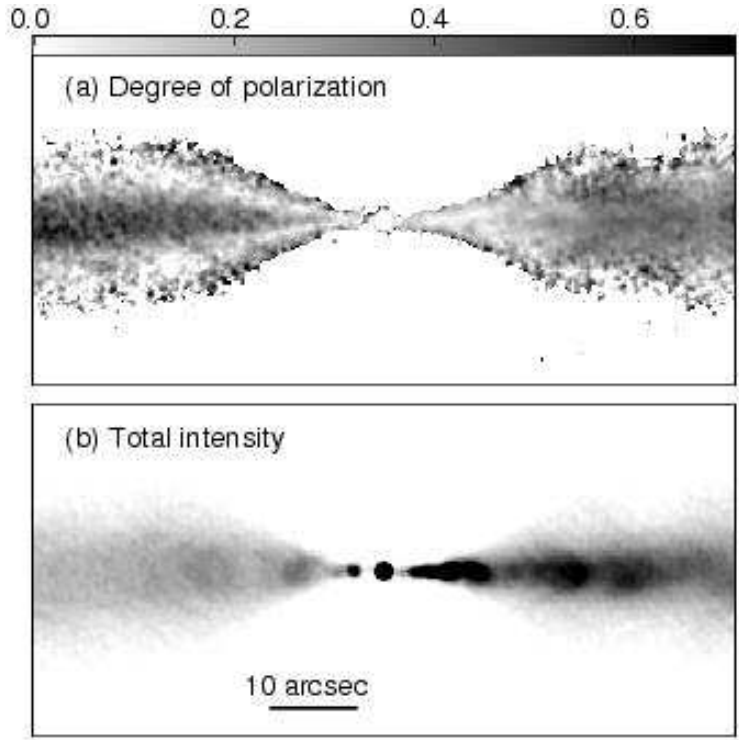}{9cm}{0}{50}{50}{-150}{-10}
\caption{Images of 3C\,296 at a resolution of 0.75\,arcsec \citep{3c296}. (a)
  degree of polarization; (b) total intensity.\label{fig:ipdata296}}
\end{figure}

\begin{figure}[!ht]
\plotfiddle{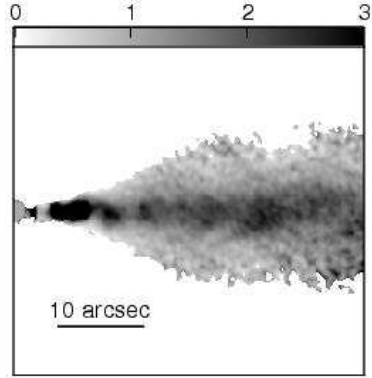}{4.5cm}{0}{25}{25}{-100}{-10}
\caption{Jet sidedness for 3C\,296, derived by dividing the total-intensity
  image from Fig.~\ref{fig:ipdata296}(b) by a copy of itself rotated by
  180$^\circ$ \citep{3c296}.\label{fig:3c296side}}
\end{figure}

An example of the characteristically asymmetric appearance of FR\,I jet bases is
shown in Fig.~\ref{fig:ipdata296}(b).  As well as this asymmetry, the main and
counter-jets also show differences in their transverse brightness profiles, in
the sense that the main jet is more centrally peaked (e.g.\
Fig.~\ref{fig:3c296side}).  This is a natural manifestation of a transverse
velocity gradient, with higher flow speeds on-axis.  There is also a systematic
difference in polarization structure between the main and counter-jets (e.g.\
Fig.~\ref{fig:ipdata296}a), in the sense that the brighter jet base has a
smaller degree of polarization on-axis with more prominent regions of
longitudinal apparent magnetic field.  If the jets are
intrinsically symmetric and relativistic, this effect must be due to aberration, which
causes the two jets to be observed at different angles to the line of sight in
the rest frames of their emitting material and therefore to have different
observed polarizations. The combination of asymmetries in total intensity and
linear polarization provides enough information to break the degeneracy between
flow velocity and angle to the line of sight, as well as to constrain the
three-dimensional structure of the magnetic field, allowing us to make detailed
models of the jets.

\section{Models of FR\,I jets: geometry, speed and field structure}
\label{models}

\begin{figure}[!ht]
\plotfiddle{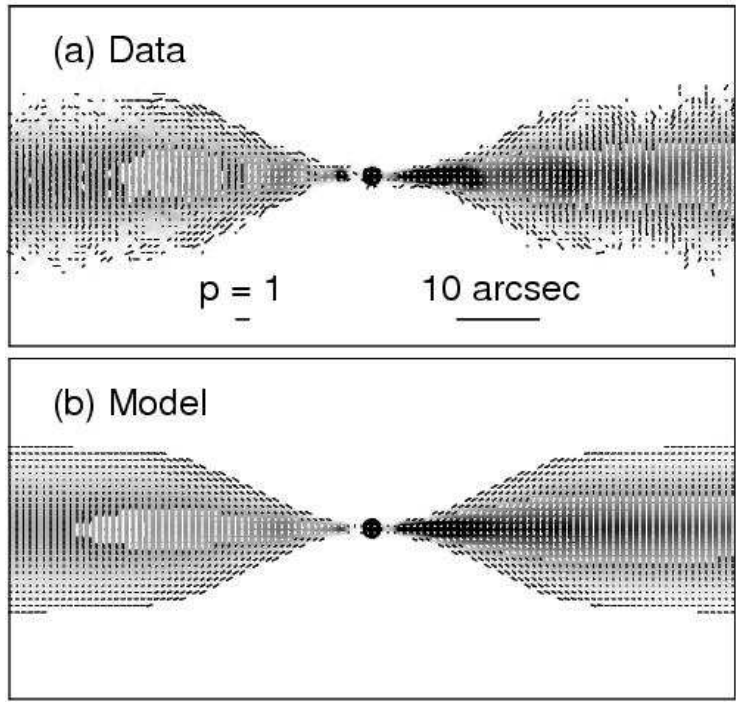}{9cm}{0}{50}{50}{-145}{-15}
\caption{A comparison between data and model for 3C\,296 \citep{3c296}. Vectors
  with magnitudes proportional to the degree of polarization, $p$ and with
  directions along the apparent magnetic field are superimposed on grey-scales
  of total intensity. (a) Data; (b) model.\label{fig:ivec296}}
\end{figure}

We have now modelled the flows in five FR\,I radio galaxies: 3C\,31,
B2\,0326+39, B2\,1553+24, NGC\,315 and 3C\,296 \citep{LB02a,CL,CLBC,3c296}.  We
assume that the jets are precisely symmetrical, axisymmetric, relativistic
flows and derive their geometries, velocity fields, emissivity and
magnetic-field component ratios by fitting to deep VLA observations.  We find that a
decelerating jet model fits the observed brightness and polarization structures
and asymmetries well (e.g.\ Fig.~\ref{fig:ivec296}).

\begin{figure}[!ht]
\plotone{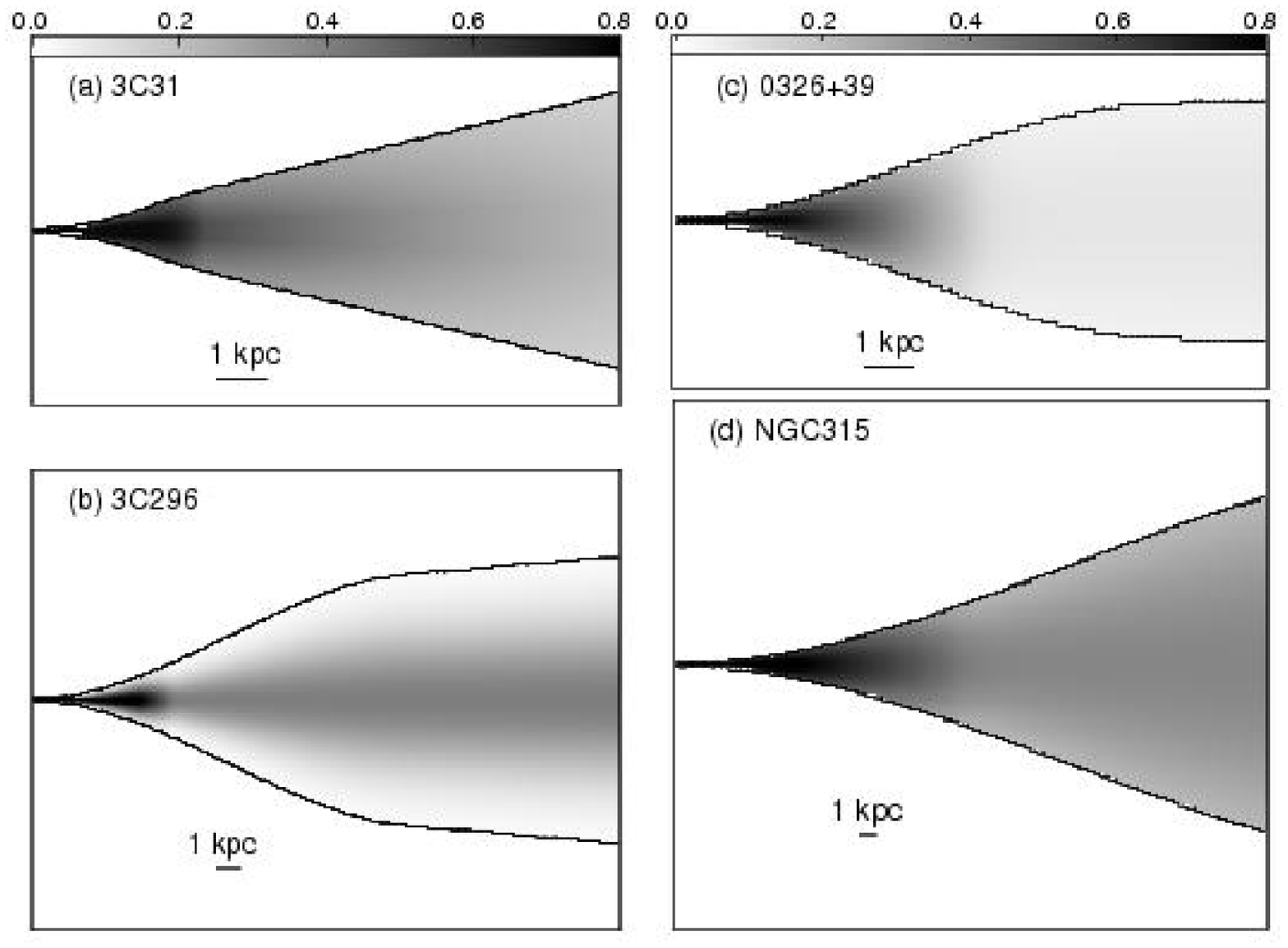}
\caption{Grey-scales of jet speed derived from relativistic models for: (a)
  3C\,31 \citep{LB02a}; (b) 3C\,296 \citep{3c296}; (c) B2\,0326+39 \citep{CL};
  (d) NGC\,315 \citep{CLBC}. The range is 0 -- 0.8$c$. \label{fig:vels}}
\end{figure}

FR\,I jets on kpc scales can generally be divided into a flaring region (where
the expansion rate first increases rapidly and then decreases again) and an
outer region in which the expansion is uniform. We cannot model the jets within
a kpc or so of the nucleus, where they tend to be faint and poorly resolved by
the VLA. Where they first brighten, we derive characteristic on-axis velocities
$\beta = v/c \approx 0.8$. Rapid deceleration then occurs over distances of 1 --
10\,kpc (always before the jets recollimate), after which the velocity either
stays constant or decreases less rapidly.  For all sources except 3C\,296, the
transverse velocity profiles are consistent with edge/on-axis ratios
$\approx$0.7 everywhere (although we cannot rule out evolution from a top-hat
profile at the initial brightening). In 3C\,296, the only source with a bridged
twin-jet structure that we have observed in sufficient detail
(Fig.~\ref{fig:fris}f), the velocity falls to a low fractional value $\la 0.1$
at the jet edge. The jets in 3C\,296 may be embedded within the lobes rather
than propagating in direct contact with the interstellar medium of the host
galaxy, as appears to be the case for the other sources.  Velocity fields for
four of the sources are shown in Fig.~\ref{fig:vels}.

\begin{figure}[!ht]
\plotfiddle{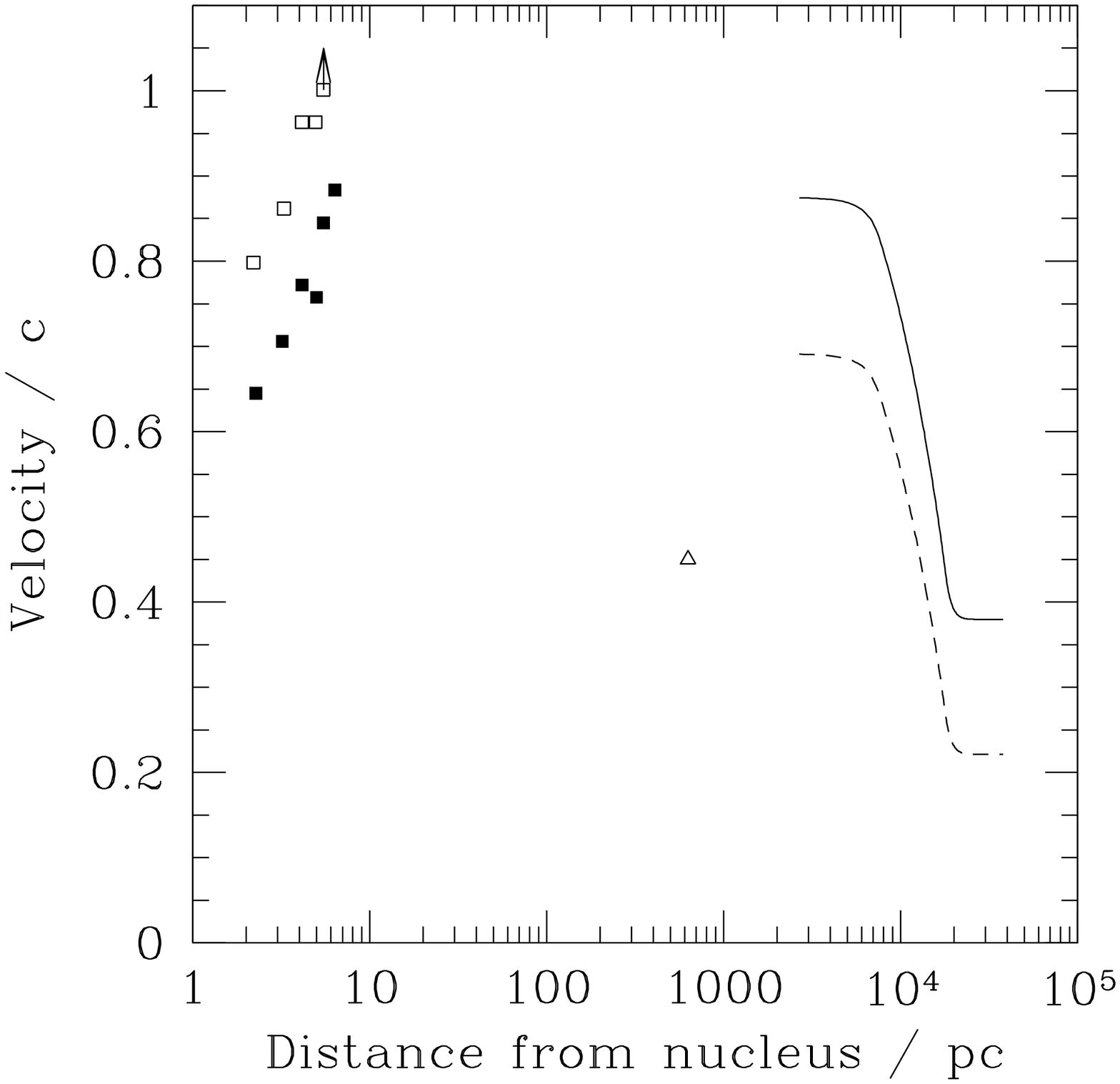}{5cm}{0}{27}{27}{-100}{-40}
\caption{A comparison of velocity estimates for NGC\,315 on pc and kpc scales,
  plotted against distance from the nucleus. Filled squares: velocities from
  proper motions; open squares: velocities from jet/counter-jet ratios (both
  from \citealt{Cotton99}); open triangle: velocity derived from the
  jet/counter-jet ratio at 0.4\,arcsec resolution close to the nucleus.  The
  full and dotted lines show model fits for the centre and edge of the jets,
  respectively \citep{CLBC}.\label{fig:accel315}}
\end{figure}

The data hint at further complexity in the jet velocity fields before the
initial brightening. Measurements of both sidedness ratio and proper motion on
pc scales in NGC\,315 indicate an apparent {\em acceleration}
\citep[Fig.~\ref{fig:accel315}]{Cotton99}. Our VLA observations also show
slightly smaller sidedness ratios in the faint, kpc-scale jet bases than at or
immediately after the brightening point for the four sources where we can
resolve the former.  If we observe a jet stratified in velocity with inadequate
transverse resolution, the speed inferred for its {\em apparently} brightest
emission will be influenced by how changes in its transverse emissivity profile
interact with the (orientation-dependent) Doppler favouritism, and may give a
misleading account of the true velocity variation along the jet.  Both the
pc-scale and faint kpc-scale regions are barely resolved in width, so we cannot
yet tell whether the bulk flow really accelerates.

\begin{figure}[!ht]
\plotone{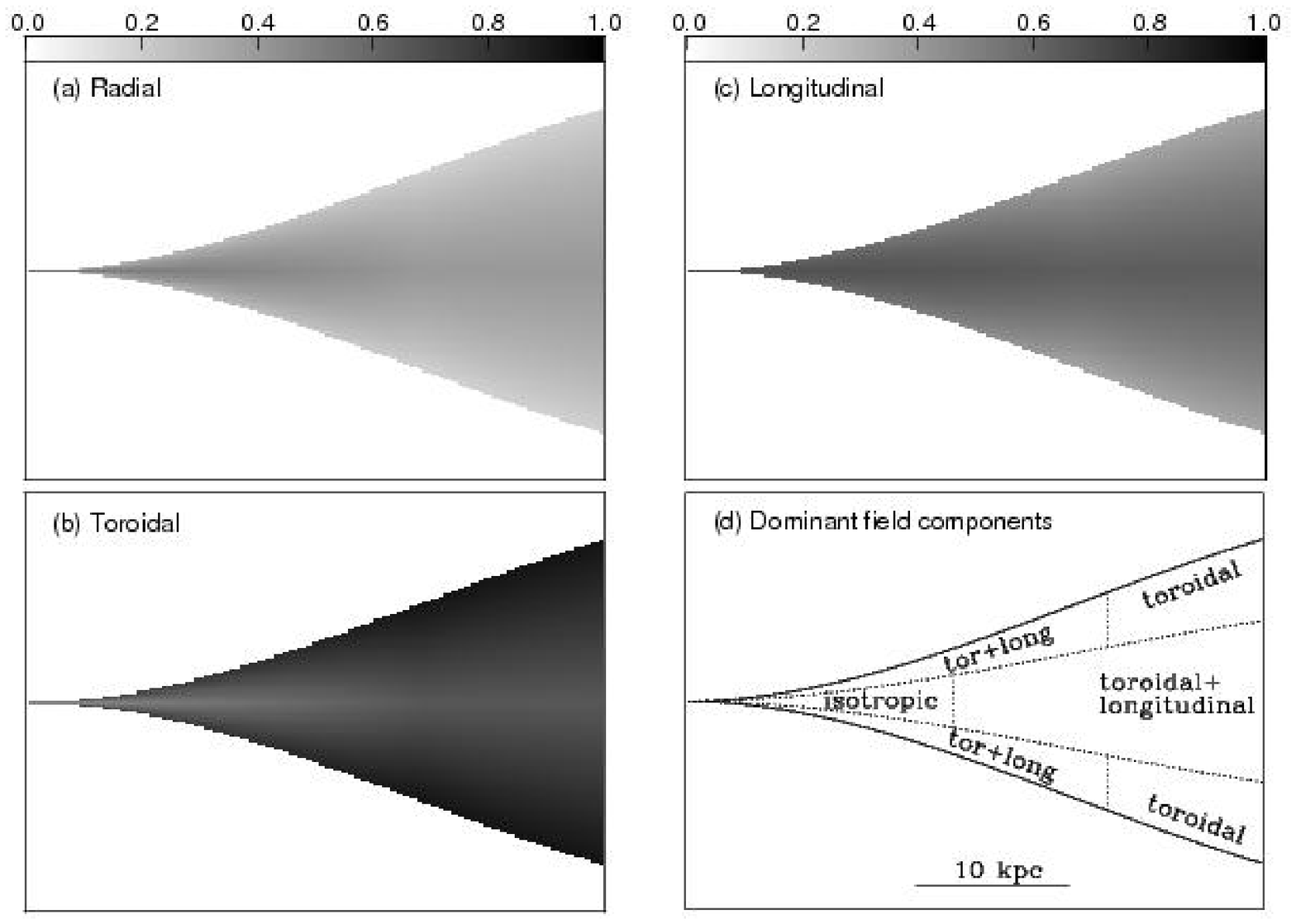}
\caption{Grey-scales showing the fractional magnetic-components in the toroidal,
  longitudinal and radial directions inferred for the jets of NGC\,315 by
  \citet{CLBC}. The dominant components are sketched in panel (d).\label{fig:bgrey}}
\end{figure}

On average, the largest single magnetic-field component is toroidal. The
longitudinal component is significant close to the nucleus but decreases with
distance and the radial component is always the smallest of the three (e.g.\
Fig.~\ref{fig:bgrey}).  We also find that the toroidal component is stronger
relative to the longitudinal component at the edge of the jet.  The evolution of
the field component ratios along the jets before they recollimate is not
consistent with flux freezing in a laminar flow, which requires a much more
rapid transition from longitudinal to transverse field than we infer. After
recollimation, flux freezing is consistent with our results, however.

Before the jets decelerate and recollimate, the
quasi-one-dimensional adiabatic approximation (together with flux freezing)
is grossly inconsistent with the emissivity evolution inferred from our models.
After deceleration and
recollimation, however, this approximation describes the emissivity
evolution along the jets quite well. 
Details of the radio spectra and high-energy emission are given by
\cite{L07}.

\section{Conservation-law analysis and jet energy fluxes}
\label{cons}

It is likely that FR\,I jets decelerate as a result of entraining matter, either
by {\it injection} of mass lost
from stars within the jet volume or 
by {\it ingestion} of ambient gas 
from the surrounding IGM via a mixing
layer. The kinematic models of Section~\ref{models} prescribe the variation of
velocity along a jet, together with its geometry.  Given the external pressure
and density profiles derived from X-ray observations, we can then apply
conservation of energy, momentum and mass in the quasi-one-dimensional
approximation, following \citet{Bick94} and including the effects of
buoyancy. Two key assumptions are needed to close the problem: the jet is
assumed to be in pressure equilibrium with the surrounding medium after
recollimation and $\Phi = \Pi c$, as expected for an initially relativistic jet
($\Phi$ is the energy flux with the rest-mass contribution subtracted and $\Pi$
is the momentum flux). In addition, a relativistic equation of state is assumed
and anisotropic magnetic stresses are neglected.  The results of this
calculation are estimates of the jet energy flux and the variations of internal
pressure, density, mass flux and entrainment rate as functions of distance from
the nucleus.

\begin{figure}[!ht]
\plotfiddle{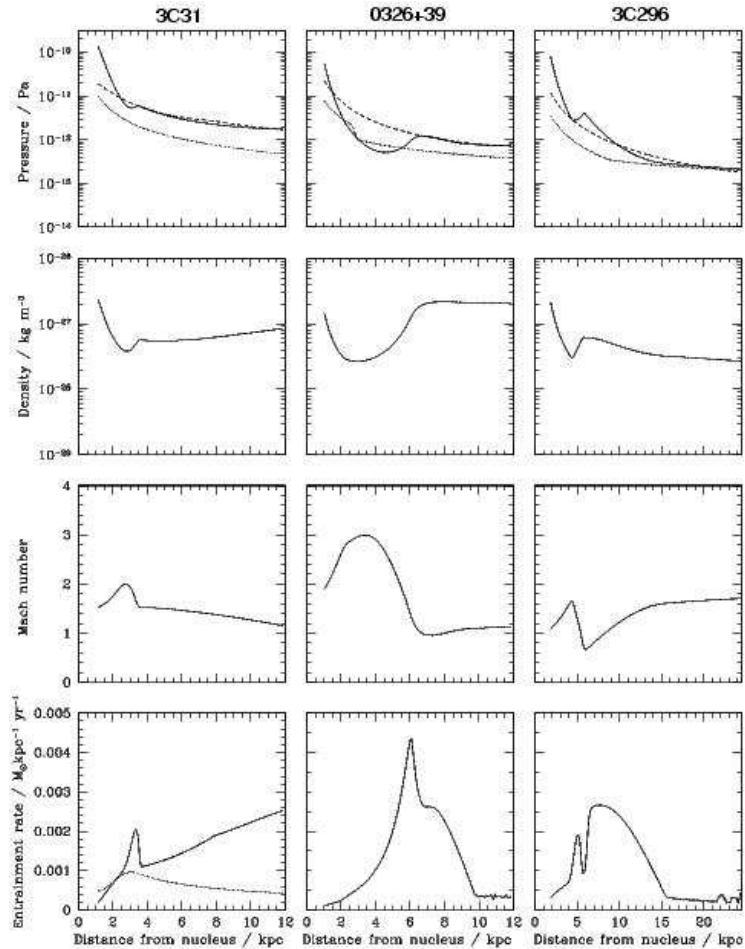}{13cm}{0}{50}{50}{-145}{-10}
\caption{Profiles of pressure, density, Mach number and entrainment rate along
  the jets of 3C\,31, B2\,0326+39 and 3C\,296 from a quasi-one-dimensional
  conservation-law analysis \citep[Laing et al., in preparation]{LB02b}. In the
  pressure plots, the full, dashed and dotted lines show the internal, external
  and synchrotron minimum pressures, respectively. The dotted curve in the
  entrainment rate plot for 3C\,31 is an estimate of mass input by injection from 
  stars within the jet volume \citep{LB02b}.\label{fig:cons}}
\end{figure}

This technique was first applied to 3C\,31 by \cite{LB02b}. In
Fig.~\ref{fig:cons} we also show the preliminary profiles of pressure, density,
Mach number and entrainment rate for two other sources -- B2\,0326+39 and
3C\,296 (Laing et al., in preparation).  The top panels show the internal,
external and synchrotron minimum pressures. By hypothesis, the first two are
matched in the outer regions, but we also note that the pressure gradients are
quite similar (for 3C\,296 it is possible that the external pressure should be
that of the lobe rather than the surrounding IGM). The synchrotron minimum
pressure is comparable with or slightly less than the internal pressure in the
outer parts, as expected if the latter is dominated by relativistic particles
and fields close to equipartition. In all three sources, there is a significant
overpressure at the brightening point (where the analysis starts) and this
drives the initial expansion in the flaring region. The pressure becomes less
than that of the surroundings at the end of the flaring region before the jet
recollimates and reaches pressure equilibrium (the large under-pressure in
B2\,0326+39 between 2 and 4\,kpc is probably due to an inaccurate estimate of
the outer velocity, which is poorly constrained).  The jet densities are very
low (typically equivalent to 1 proton m$^{-3}$), and the initial density ratios
with the surroundings are $\sim$10$^{-5}$.  The jets are transonic, with Mach
numbers $\sim$1.  The entrainment rate for 3C\,31 continues to rise
at large distances whereas the profiles for B2\,0326+39 and 3C\,296 both show
peaks in the flaring region before dropping to very low values. The reason is
that the kinematic model for 3C\,31 requires continued deceleration after
recollimation, whereas those for the other two sources indicate an approximately
constant asymptotic velocity.   Differences in external density and/or shielding
by lobe material may be responsible. The entrainment rate profile for 3C\,31
also shows a prediction for the mass input from stars in the jet volume
\citep{LB02b}. This is comparable with the required entrainment rate close to
the start of the model, but not at large distances, suggesting that 
mass injection and ingestion are both important.

The derived energy fluxes are $\Phi \approx 1.1 \times 10^{37}$, $7 \times
10^{36}$ and $1.6 \times 10^{36}$\,W, respectively, for 3C\,31, 3C\,296 and
B2\,0326+39.  It will be of interest to compare these estimates with values
derived from cavity dynamics \citep[and references therein]{McN07}.
Estimates of the mass density and the number density of radiating electrons can
be combined to constrain the composition of the jets (see \citealt{LB02b} for a
detailed review of assumptions). At the brightening point, limiting cases
include:
\begin{enumerate}
\item A power-law Lorentz-factor distribution $n(\gamma)d\gamma \propto
  \gamma^{-2.2}d\gamma$ of relativistic electrons, each associated with
  a (cold) proton. In this case, there must be a low-energy cut-off at
  $\gamma_{\rm min} \approx 35$ (3C\,31 and B2\,0326+39) or $\approx 300$
  (3C\,296).
\item An electron-positron jet with entrained thermal matter providing
  essentially all of the mass.
\end{enumerate}
Unfortunately, it is not yet possible to distinguish between these (and other)
compositions. 

\section{Where next?}

The techniques reviewed here have led to a detailed picture of the evolution of
the FR\,I jets as they flare and decelerate. They work because we can compare
both jets in the same source: sizes match the resolution of our most
sensitive instruments, surface brightnesses are (just) high enough, Faraday
rotation is easy to correct and jet speeds are modest. The key project when
EVLA and eMERLIN become operational will be an attempt to
apply the same methods to other types of relativistic jet: closer to the
launching scale and in FR\,II sources or microquasars.  There are also
more immediate questions, including:
\begin{enumerate}
\item What makes FR\,I jets brighten so abruptly in the radio band at distances
  of a few kpc from the nucleus?
\item Can we decide what combination of stellar mass-loss (injection) and 
external entrainment (ingestion) operates in FR\,I jets?
\item Can we (finally) detect internal depolarization and hence estimate
  densities of thermal matter now that we
  understand foreground Faraday screens much better?
\item Do different methods for estimating FR\,I jet powers (conservation-law,
  cavity dynamics, bow-shock physics) give the same answer? Can they be made consistent
  with radiation-loss timescales?
\item Do jets really accelerate on pc scales and/or as they flare or is this
effect an artefact of inadequate transverse resolution?
\end{enumerate} 
\acknowledgements 
The National Radio Astronomy Observatory is a
facility of the National Science Foundation operated under cooperative agreement
by Associated Universities, Inc.


\end{document}